\documentclass{atmp}
\usepackage{amscd,amssymb,amsthm,amsmath,graphicx,psfrag}
\usepackage[all]{xy}

\newtheorem{theorem}{Theorem}[section]
\newtheorem{lemma}{Lemma}[section]
\newtheorem{definition}{Definition}[section]

\newcommand{\A}{\ensuremath{\mathcal{A}}}

\newcommand{\KC}{\ensuremath{{\mathbb C}}}
\newcommand{\KR}{\ensuremath{{\mathbb R}}}
\newcommand{\KZ}{\ensuremath{{\mathbb Z}}}
\newcommand{\KCP}{\ensuremath{{\mathbb C}{\mathbb P}}}
\newcommand{\KT}{\ensuremath{{\mathbb T}}}

\newcommand{\Cpct}{\ensuremath{\mathcal{K}}}

\newcommand{\CrPr}[3]{ \ensuremath{#1 \underset{#3}{\rtimes} #2} }

\numberwithin{equation}{section}

\renewcommand{\arxurl}[1] {%
  \barefootnote{
    $^\dag$Current address:
Theoretical Physics, RSPhysSE, ANU, Canberra, ACT 2602.\\
    {\small e-print archive:}~{\texttt http://lanl.arXiv.org/abs/#1}}
}

\graphicspath{{H:/IP/ATMP/12-1/Pagination/Art/}}

\copyrightnotice{2007}{12}{185}{215}

\begin{document}

\title[Topological T-duality and KK-monopoles]{Topological
T-duality and Kaluza--Klein monopoles}

\arxurl{arXiv:math-ph/0612034}

\author[Ashwin S. Pande]{Ashwin S. Pande$^\dag$}

\address{Department of Mathematics, University of Maryland at
College Park, College Park, MD 20742, USA\\
asp105@rsphysse.anu.edu.au}

\begin{abstract}
We study topological T-duality for spaces with a semi-free
$S^1$-action with isolated fixed points. Physically, these
correspond to spacetimes containing Kaluza--Klein monopoles. We
demonstrate that the physical dyonic coordinate of such spaces
has an analogue in our formalism. By analogy with the Dirac
monopole, we study these spaces as gerbes. We study the effect
of topological T-duality on these gerbes.
\end{abstract}

\maketitle

\section{Introduction\label{SecIntro}}
String theory replaces point particles by extended objects
called strings propagating in a background spacetime called the
``Target Space'' \cite{Polchinski}.  Torus duality or T-duality
in string theory is an important symmetry of string theories in
which the target space has a torus action: IIA string theory
with a target spacetime $X$ possessing a $T^n$-action is
identical to IIB string theory on the T-dual spacetime $X^{\#}$
\cite{Alvarez}; i.e., the two quantum theories are identical up
to a canonical transformation.

\pagebreak

In the case that spacetime is a principal $T^n$ bundle over a base,
the dual spacetime $X^{\#}$ is roughly one in which each torus orbit
in the original spacetime has been replaced by its dual torus
(following Ref. \cite{MRCMP}, if $\Lambda$ is a lattice in $\KR^n$
and $\Lambda^{\ast}$ is the dual lattice in the dual vector space
$(\KR^n)^{\ast},$ then the torus dual to $\KR^n/\Lambda$ is
$(\KR^n)^{\ast}/{\Lambda^{\ast}}$). In the original string theory
calculation, the T-dual is obtained as a manifold together with
various extra data (metric, B-field, dilaton, and R-R
charges).\footnote{It was the study of the mapping of the R-R charges
of string theory under such a transformation that gave the initial
impetus to study T-duality purely topologically \cite{BouEvMa}.}

Surprisingly, if we restrict ourselves to spacetimes which are
principal $T^n$ ($n=1,2$) bundles over some base, it is possible
to develop a theory of T-duality using only topological
information.\footnote{The metric structure and finer structures
(like the Kahler structure, needed for supersymmetry) are not
needed to determine the topological type of the T-dual.}That is,
it is possible to specify the topological structure of the
T-dual of a given spacetime using only the $H$-flux and
topological structure of the original spacetime.\footnote{Here
``topological structure'' should be interpreted broadly because
the T-dual may be a non-commutative space.}This is surprising
since string theory usually requires a smooth, semi-Riemannian
manifold (usually a K\"ahler manifold) as its target. The
resulting theory has been the subject of the papers [3, 4, 5] and may be viewed as a ``topological
approximation" to T-duality in string theory. That is, it should
be possible to take a spacetime $X$ which possesses a T-dual in
the sense of Ref. \cite{MRCMP} and give it additional structure
(metric, spin etc.) so as to construct on $X$ a type IIA string
theory and to construct on $X^{\#}$ a type IIB string theory
which form a dual pair (however, see Ref. \cite{BEMPRL}).

Suppose $X$ is homeomorphic to a smooth closed, connected
Riemannian manifold. Suppose, further that $X$ possesses a
smooth, semi-free, action of $T^n, (n=1,2).$ (By a semi-free
action, we mean an action with exactly two types of orbits: free
orbits and fixed points.) Then we are in the basic setup of Ref.
\cite{MRCMP} if a class $\delta \in H^3(X,\KZ)$ is specified.
Here $X$ is to be viewed as a spacetime and the map $X\to X/T^n$
is a degenerate fibration.

If the semi-free action has no fixed points, it is free and so
by the Gleason slice theorem, $X$ is a principal $T^n$-bundle
over $X/T^n.$ This case has already been extensively studied in
Refs. \cite{MRCMP,RaeRos}. There, the authors associate to a
space $X$ with $H$-flux $\delta$, the continuous trace algebra
CT$(X,\delta).$ If $X$ is a principal $T^n$-bundle ($n=1,2$),
they demonstrate the following:
\begin{itemize}
\item If $n=1,$ there is a unique lifting $\alpha$ of the $\KR$-action on $X$ to
CT$(X,\delta).$ The T-dual spacetime to $X$ is given by the
spectrum of CT$(X,\delta)\underset{\alpha}{\rtimes} \KR.$ In
this case, the crossed product is always continuous trace.

\item If $n=2$ and if a certain condition is satisfied, the T-dual
is unique and is given by the spectrum of the crossed product as
above. If this condition is violated, however, there is no
unique lifting of the $\KR^n$-action on $X$ to CT$(X,\delta).$
Also, the crossed product is not Type I. However, as explained
in Ref. \cite{MRCMP}, the T-dual may be viewed as a
non-commutative space.
\end{itemize}
In either case, the natural action of $\hat{\KR}^n$ on the
spectrum of the crossed product makes it into a principal
$T^n$-bundle over $X/\KR^n.$

In this paper, we study the T-duals of semi-free $S^1$-spaces.
In Section~\ref{SecKKTdual}, we show that the formalism of Ref.
\cite{MRCMP} extends to spaces containing KK-monopoles. In
Section~\ref{SecDyon}, we show that the result obtained in the
previous section may be used to define an analogue of the dyonic
coordinates of a system of KK-monopoles within the formalism of
topological T-duality. In Section~\ref{SecGerbe}, we study the
result of Section~\ref{SecKKTdual} using gerbes.

\section{Kaluza--Klein monopoles and T-duality\label{SecKKTdual}}

In this paper, we study semi-free $S^1$-spaces with non-empty
fixed point sets. Now, it is well known \cite{Bredon} that a
smooth action of a compact Lie group $G$ on a smooth manifold
$M$ in a neighborhood of a fixed point is
equivariantly-$G$-homeomorphic to an orthogonal action of $G$ on
a finite-dimensional vector space. It would be helpful to
consider this case first.

\enlargethispage{2pt}

Thus, now we let $X$ be $\KR^k$ with a faithful orthogonal action of
$S^1.$ We view this as defining a fibration of $\KR^k$ over the
quotient space. We attempt to define a T-dual for such a fibration.
Note that there can be no $H$-flux in this setting, since $\KR^k$ is
contractible (i.e., $\delta = 0$). As in Ref. \cite{MRCMP}, we
introduce the $C^{\ast}$-algebra ${\mathcal A} =
\hbox{CT}(X,\delta)$. We lift the $S^1$-action on $X$ to a
$\KR^n$-action $\alpha$ on $\mathcal A$. This may always be done for
all $n$ because $X$ is contractible (see Raeburn and Rosenberg
\cite{RaeRos}.)\footnote{For other spaces, if $n>1$, and
$H^3(X,\KZ)\neq 0$ Mackey obstructions may arise which prevent this
lifting.} As in Ref.~\cite{MRCMP}, we may attempt to define the T-dual
of $X$ as the spectrum of the crossed product ${\mathcal{A}}
\underset{\alpha}{\rtimes} \KR$.

As a test example, let $X=\KC^2$ with the $S^1$-action
\[
\hbox{e}^{2\pi {\rm i} \theta} \cdot (z,w)= (\hbox{e}^{2 \pi {\rm i} \theta}
z, \hbox{e}^{2 \pi {\rm i} \theta} w).
\]
This may be lifted to the obvious action of $\KR$ namely,
\[
\alpha_t(z,w)= (\hbox{e}^{2\pi {\rm i} t}z, \hbox{e}^{2 \pi {\rm i} t} w).
\]
Note that the $S^1$-action leaves each 3-sphere
\[
S^3_r=\{(z_1,z_2) \mid |z_1|^2 + |z_2|^2 = r^2\}
\]
invariant. The $S^1$-action on each $S^3_r$ has $S^2$ as quotient. The
quotient map is the Hopf fibration. The origin $(0,0)$ is a
fixed point for the $\KR$-action.

We may also view this as a fibration of  $C^0S^3$ over the base
$C^0S^2$ (here, $C^0S^3 \simeq (S^3\times\KR^{+})/(S^3\times{0})$)
the open cone on $S^3$). The map sending $C^0S^3$ to $C^0S^2$ is the
Hopf fibration when restricted to $S^2 \times \{t\}, t\neq 0$ and
sends the vertex of $C^0S^3$ to the vertex of $C^0S^2.$ (Note
$S^3\times (0,\infty)$ is a principal $S^1$ bundle over $S^2 \times
(0,\infty).$)\vspace*{-3pt}

\subsection{Physical T-duals}

As we mentioned in the introduction, T-duality was first
discovered in the theory of closed strings. This formalism only
allows us to calculate the\break T-dual of space with a free
$T^n$-action. It was soon realized that string theory contains a
theory of extended objects called ``branes''. These are (roughly)
submanifolds of spacetime on which strings can end. Due to the
strings ending on such a submanifold,\footnote{The submanifold
associated with a brane is termed its ``worldvolume''.} there is a
quantum field theory defined on it. There are two types of
branes: the D$p$-branes are submanifolds which are sources of
the R-R fields of string theory;\footnote{D$p$-branes possess
$(p+1)$-dimensional worldvolumes. In type IIA string theory, $p$
can only be even dimensional. In type IIB string theory, $p$ can
only be odd dimensional. Their worldvolume theory at low
energies is a Super-Yang--Mills gauge theory.}the NS5-branes,
on the other hand are submanifolds which are sources of the
Neveau-Schwartz-B-field.\footnote{These have six-dimensional
worldvolumes. Their worldvolume theory is a string theory (a
``little string theory''). At low string coupling, they are
extremely massive compared to D$p$-branes.}After the
introduction of D$p$-branes, T-duality was studied by putting
D$p$-brane probes into a geometry. The T-dual is the moduli space
of the worldvolume theory on the D$p$-brane. This approach is
extremely flexible and enables the calculation of several
T-duals unobtainable by previous methods.

It is useful to study the T-duals of semi-free $S^1$-spaces in
the physics literature.
\begin{enumerate}
\item[1)] The T-dual of a NS5-brane is a Kaluza--Klein (KK) monopole (see
Refs. \cite{NSDual,UrWeb,Sorkin}). Geometrically, an NS5-brane
is a six-dimensional submanifold of $X$ which is a source of
$H$-flux. Topologically, $X$ is a fibration $\KR^6 \times
C^0S^2\times S^1 \overset{\pi}{\to} \KR^6\times C^0S^2$, where
$\pi$ is the projection map. The worldvolume of the NS5-brane
intersects the $S^1$-fiber $\pi^{-1}(0)$ at a single point while
its six worldvolume directions occupy $\KR^6$. Since $\KR^6$ is
contractible,  it does not affect the topological type of the
T-dual. In the following, we will model this by studying
$C^0S^2\times S^1 \overset{\pi}{\to} C^0S^2.$ We will say that
the NS5-brane is sitting at some location on the $S^1$-fiber
over $0.$ This brane emits $1$ unit of $H$-flux which we model
as the cohomology class $[1]\in H^3(CS^2\times S^1 - \{0\}
\times S^1)$, where $[1]$ is the generator of
\[
\hskip2.2pc H^3(S^2\times S^1)\simeq H^3(S^2\times (0,\infty ) \times S^1)
\simeq H^3(C^0S^2\times S^1 - \pi^{-1}(0)).
\]

A Kaluza--Klein monopole is a semi-Riemannian
manifold,\footnote{see Refs. \cite{Sorkin,Jens,Harv}.}which
solves Einstein's equations. Topologically, this manifold is $\KR^5$
with metric $g_{\rm KK} = -dt^2 + g_{\rm TN}.$ Here $g_{\rm TN}$ is a
certain Riemannian metric on $\KR^4$ called the Taub-NUT
metric.\footnote{This is an example of an ALF gravitational
instanton metric.}The space $(\KR^4,g_{\rm TN})$ (termed
``Taub-NUT'' space) possesses an isometric action of $S^1$ with
one fixed point, and is $S^1$-equivariantly homeomorphic to
$\KC^2$ with the $S^1$-action $\lambda \cdot (z_1,z_2) =
(\lambda \cdot z_1, \lambda \cdot z_2), \lambda \in S^1
\subseteq \KC, z_i \in \KC.$ In the  physics literature, the
time direction is often ignored and $(\KR^4, g_{\rm TN})$ is also
called the ``Kaluza-Klein monopole''. For our purposes, a
Kaluza--Klein monopole is an $S^1$ fibration over $\KR^3 \simeq
C^0S^2.$ Over each sphere $\{t\} \times S^2 $ in the base, the
fibration is the Hopf fibration. Over $0$ in the base, the
fibration degenerates to a point. It may be viewed as a
fibration $C^0S^3 \overset{\pi^{\#}}{\to} C^0S^2.$

Thus, we say that the string-theoretic T-dual of $C^0S^3
\overset{\pi^{\#}}{\to} C^0S^2$ is $C^0S^2\times S^1
\overset{\pi}{\to} C^0S^2$ together with the $H$-flux emitted
from a point source sitting at some point on the fiber
$\pi^{-1}(0)$ (also termed a\break $H$-monopole).  $\pi^{\#}: C^0S^3
\to C^0 S^2$ is the test case discussed in Section~\ref{TestEx}. It
is an important example, because (as will be seen below) most
physical examples are built up from this one.

This T-dual may be calculated using Buscher's rules. We use
polar coordinates $(r,\theta,\phi)$ to parametrize the base and
a periodic coordinate $\kappa$ to parametrize the $S^1$-fiber.
Following Ref.~\cite{Jens}, the Riemannian metric on Taub-NUT
space may be written as
\begin{equation} \label{gTaubNUT}
g_{\rm TN} = H(r) d\vec{r} \cdot d\vec{r} + H(r)^{-1}\left(d\kappa +
\frac{1}{2} \vec{\omega}\cdot d\vec{r}\right)^2,
\end{equation}
where $d\vec{r} = (dr, d\theta,  d\phi),$ (note that $\phi$ is the
periodic polar coordinate). $H(r) = g^{-2} + (2r)^{-1}$, and
$\omega_r=\omega_{\theta}=0, \omega_{\phi} = (1-\cos(\theta)).$ We
may\break write
\begin{align}\label{EqTNMetric}
g_{\rm TN}&=Hdr^2+Hr^2\sin^{2}(\theta)(d \phi)^2+Hr^2(d\theta)^2
+H^{-1}(d\kappa)^2 \nonumber \\
& \quad{} +H^{-1}(1-\cos(\theta))d\phi d\kappa +
H^{-1}(1-\cos(\theta))^2(d\phi)^2,
\end{align}
where $\kappa$ is the coordinate on the $S^1$-fiber. Let
$x^0=\kappa,x^1=r,x^2=\theta,x^3=\phi.$ Buscher's rules specify the
T-dual metric ($\tilde{g}$) and $B$-field ($\tilde{b}$) in terms of
the original metric ($g$) and $B$-field ($b$). By Ref.
\cite{Alvarez},\break we~have\vspace*{1.5pt}
\begin{align}\label{EqBuscher}
&\quad \ \tilde{g}_{00} = 1/g_{00}, \tilde{g}_{0\alpha}=b_{0\alpha}/g_{00},\nonumber\\
&\tilde{g}_{\alpha \beta}= g_{\alpha \beta} -
(g_{0\alpha}g_{0\beta}- b_{0\alpha}b_{0\beta})/g_{00}.
\end{align}
Now, we have $b_{\alpha \beta} = 0$, $g_{00}= H$, $g_{0 \alpha}=0$ by
equation~(\ref{EqTNMetric}) above. Therefore $\tilde{g}_{00} = H,
\tilde{g}_{0\alpha}=0\, \forall \alpha.$ Also by
equation~(\ref{EqBuscher}), $\tilde{g}_{\alpha \beta} = g_{\alpha
\beta}$ if $\alpha \neq 3$ or $\beta \neq 3.$ We also
have\vspace*{1.5pt}
\begin{align}
\tilde{g}_{33} & = g_{33} - (g_{03})^2/(H^{-1}) \nonumber \\
& = Hr^2 + H^{-1}(1-\cos(\theta))^2
-H^{-1}(1-\cos(\theta))^2=Hr^2. \nonumber
\end{align}
Hence,\vspace*{1.5pt}
\begin{align}\label{gHMono}
g_{H} &
=H(d\kappa)^2+H(dr)^2+Hr^2\sin^2(\theta)(d\phi)^2+Hr^2(d\theta)^2
\nonumber \\
& =H((d\kappa)^2+ d\vec{r} \cdot d\vec{r}).
\end{align}
It is clear that $\tilde{g}$ is conformally equivalent to a
product metric on $\KR^3\times S^1.$ As $r\to 0, H\to \infty$
thus implying that the $S^1$-fiber over $0\in \KR^3$ is
infinitely far away from the rest of the space. This is termed
as a \textit{smeared} $H$-monopole solution. Quantum effects
\cite{Jens} are supposed to modify $H$ so that $\lim_{r\to 0}\,
H(r,\theta)$ is finite except at the value of $\theta$
corresponding to the location of the $H$-monopole.

\item[2)] The T-dual of a set of $p$ distinct, non-intersecting NS5-branes
is a \hbox{$p$-center} KK-monopole (see Ref.~\cite{NSDual}). This is
obtained from the previous example in the obvious fashion: we
introduce $p$ sources of $H$-flux in spacetime for the $p$
NS5-branes. In the T-dual, we allow the $S^1$-fiber to
degenerate to a point over $p$ points in the base.

Let $(X_p,g_p)$ be the spacetime containing $p$ KK-monopoles.\footnote{This is termed a multi-Taub-NUT space.}This space
posseses an isometric action of $S^1$ which is free except for
$p$ fixed points. The quotient of $X_p$ by the $S^1$-action is
$\KR^3$ with the Euclidean metric. Let $\pi:X_p \to \KR^3$ be
the quotient map. If we use polar coordinates $(r,\theta,\phi)$
on the base $\KR^3$, then we may write the metric $g_p$ in terms
of these three coordinates and an additional coordinate $\kappa$
on the $S^1$-fiber. Let
\[
H({\mathbf r}) = g^{-2} + \sum_{i=1}^{p} \frac{1}{ 2|{\mathbf r}
- {\mathbf r}_i|},
\]


\noindent then,
\begin{equation}\label{EqMultKK}
g_p = H({\mathbf r}) d\vec{r} \cdot d\vec{r} + H({\mathbf
r})^{-1}\left(d\kappa +
\frac{1}{2} \vec{\omega}\cdot d\vec{r}\right)^2,
\end{equation}
$\nabla H = \nabla \times \vec{\omega}.$ The above expression for
$g_p$ agrees with the expression for $g_{\rm TN}$ above (see
equation~(\ref{gTaubNUT})) except that the expression for $H$ is
different in the two cases. Since the expression for $H$ does not
appear in the application of Buscher's rules above, the form of the
metric on the T-dual is the same in both cases. Therefore, the metric
on the T-dual of $g_{p}$ is
\begin{equation}
\tilde{g}= H\left((d\kappa)^2+ d\vec{r} \cdot d\vec{r}\right). \nonumber
\end{equation}
This is conformally equivalent to a product metric on $\KR^3
\times S^1.$\break The~fibers over $r_i, i= 1,\ldots,p$ are infinitely
far away from the rest of the spacetime, similar to the T-dual
of one KK-monopole.

Now, consider the case $p=2$. Let $Y$ be a line segment joining
the image of the two centers in $\KR^3.$ We have $\KR^3-Y \simeq
S^2\times (0,\infty).$ Let $W = \pi^{-1}(Y)$  and consider $X_2
- W.$ Now, $\forall t, \pi^{-1}(S^2\times \{t\})$ is
homeomorphic to $S^3/\KZ_2$ since each $S^2$ encloses $Y$ in
$\KR^3.$ Therefore,\footnote{i.e., the map $\pi$ restricted to
any $\pi^{-1}(S^2) \subset (X_2 - W)$ will be the projection map
of the $S^1$-bundle over $S^2$ of Chern class $2$.}$X_2 - W
\simeq (S^3/\KZ_2) \times (0,\infty)$ (this may also be seen by
examining the expression for $g_2$). Suppose we collapse $Y$ to
a point in $X_2$ and $\pi(Y)$ to a point in $\KR^3$, we would
obtain an equivariant fibration $\tilde{\pi}:C(S^3/\KZ_2) \to
C(S^2)$. Note that both $C(S^3/\KZ_2) \simeq X_2/W$ and $C(S^2)
\simeq \KR^3/W$ are contractible to their vertices. This implies
that $X_2$ is homotopy equivalent to $W = \pi^{-1}(Y) \simeq
S^2$ (in fact equivariantly so).

If $p>2,$ we may always change the total space by a
homeomorphism so that the image of the $p$ centers in $\KR^3$
under $\pi$ lie on a straight line $W.$ The inverse image of $W$
under $\pi$ is a collection of $(p-1)$ spheres joined to each
other at one point and is homeomorphic to a wedge of $(p-1)$
spheres. By an exactly similar argument to the above, $X_p$ is
homotopy equivalent to this wedge of $(p-1)$ spheres.

\item[3)] The T-dual of a $H$-monopole of charge $p$ is a KK-monopole of charge
$p$.\footnote{A space with a KK-monopole of charge $p > 1$ is not a
smooth manifold, it possesses a conical singularity at the location
of the monopole.}A $H$-monopole of charge $p$ is a fibration of the
form $C^0S^2\times S^1 \overset{\pi}{\to} C^0S^2.$ The $H$-monopole
sits at some point in $\pi^{-1}(0)$ and emits $p$ units of $H$-flux
on $C^0S^2-\pi^{-1}(0).$ We represent the $H$-flux as the cohomology
class $[p]\in H^3(\hbox{CS}^2-\{0\}\times S^1) \simeq H^3(S^2\times
S^1).$ The KK-monopole of charge $p$ is similar to the KK-monopole
configuration above, except that the fiber over each $S^2\times\{t\}$
in the base is a $S^1$ bundle of Chern class $p,$ i.e., we have a
fibration like ${CL}(1,p) \to {CS}^2$ where $L(1,p)\to S^2$ is the
lens space viewed as a principal $S^1$-bundle.\footnote{Note that
$\hbox{CL}(1,p) \simeq \KC^2/\KZ_p$, where $\KZ_p \subseteq
\hbox{SU}(2)$ in its fundamental representation.}\vspace*{-1pc}
\end{enumerate}

\subsection{Test Example\label{TestEx}}

To the test example, we will associate ${\mathcal{A}} \simeq
\hbox{CT}(C^0S^3,0),$ as there is no $H$-flux on $\KC^2.$ By
Ref.~\cite{RaeRos}, the $\KR$-action on $X \simeq C^0S^3$ lifts to a
unique $\KR$-action $\alpha$ on $\mathcal{A}.$ We work with the
example of Section~\ref{SecKKTdual}. Suppose we consider
CT$(X,0)$, then, as shown in \cite{RaeRos}, the $S^1$-action on
$X$ lifts uniquely to a $\KR$-action on $X$. By Theorem~4.8 of
Ref. \cite{RaeRos}, the spectrum of the crossed product is
homeomorphic to $(X\times \hat{\KR})/\sim$, where $\sim$ is the
equivalence relation given by
\[
(x,\gamma) \sim (y,\chi) \Leftrightarrow \overline{\KR\cdot x} =
\overline{\KR \cdot y}
\]
and $\gamma \bar{\chi} \in
(\protect{\mbox{Stab}}_x)^{\perp}.$ For $\KR,$ all irreps are
one-dimensional, and of the form $\pi_k : x \mapsto \hbox{e}^{{\rm i}kx}.$
Distinct values of $k$ correspond to non-unitarily-equivalent
irreps. (As a topological space ${\KR}\hat{}$  is homeomorphic
to $\KR.$) If $\gamma = k_1$ and $\chi = k_2$, then $\gamma
\bar{\chi}$ corresponds to $x \mapsto \hbox{e}^{{\rm i}(k_1-k_2)x}.$ We have
$\gamma \bar{\chi} \in (\protect{\hbox{Stab}}_x)^{\perp}$ iff
$\gamma \bar{\chi}(l) = 1, \, \forall l \in
(\protect{\mbox{Stab}}_x).$ If $x \neq 0,$ the stabilizer is
$\KZ,$ so, we have $\hbox{e}^{{\rm i}(k_1-k_2)n} = 1, \forall n \in \KZ.$
This implies that $(k_1 - k_2 ) = 2 l \pi,\, \forall l \in \KZ.$
Thus, points in $\hat{\KR}$ are periodically identified.
Further, points in the same $S^1$-orbit are identified. So, for
$x\neq 0,$ we have the dual principal bundle $S^2 \times S^1$ as
described in Ref. \cite{MRCMP}. If $x=0,$ the stabilizer is
$\KR,$ so, we have $\hbox{e}^{{\rm i}(k_1-k_2)x} = 1, \forall x \in \KR.$
This implies that $k_1 - k_2 = 0.$  Thus, at the fixed point
there is no quotienting.

Pick a $S^1$-invariant neighborhood $U$ of $0\in\KC^2,$ and an
$\epsilon$-neighborhood $V_i^{\epsilon}$ of $k_i.$ Then,
$W_i^\epsilon= U\times V_i^{\epsilon}$ is a neighborhood of
$\{0\}\times k_i$ in $X\times \hat{\KR}.$ The $W_i^{\epsilon}$
form a neighborhood base $k_i$ in $X\times \hat{\KR}.$ Note
that the quotient map associated to $\sim$ is open. The
saturation of $W_i^{\epsilon}$ with respect to $\sim$ is
\[
\tilde{W_i^{\epsilon}} = U \times \left(\coprod_j V_i + 2 \pi j\right)
\]
Thus, if $k_i-k_j \neq 2 \pi l, l \in \KZ,$ $\tilde{W_i}$ can be
chosen to be disjoint from $\tilde{W_j}$ by taking $V_i$ small
enough. Conversely, if $k_i-k_j = 2 \pi l,$ it is impossible to
choose disjoint neighborhoods for them in $X\times \hat{\KR}$
(since the $\tilde{W_i}$ form a neighborhood basis at $k_i$).

So, we see that the crossed product has a very non-Hausdorff
spectrum. In particular, its spectrum is $S^2  \times S^1 \times
(0,\infty)$ with the line $\hat{\KR}$ glued on at $0$. The
gluing is such that if a sequence $\{x_i\} \in S^2 \times S^1
\times (0,\infty)$ converges to $x_{\infty} \in (0 \times
\hat{\KR}) ,$ then it converges to $x_{\infty}+2\pi l, \forall l
\in \hat{\KR}.$ Note that if we remove the fixed point of the
$\KR$-action on $X$, the crossed product is a non-trivial
continuous-trace algebra CT$(S^2\times S^1 \times
(0,\infty),\delta^{\#})$.

The crossed product $\mathcal{A} \underset{\alpha}{\rtimes} \KR$
is not continuous trace, since $\hat{\mathcal{A}}$ is not $T_2$.
However, it may be viewed as a $C^{\ast}$-bundle over $C^0S^2
\times S^1$. Restricted to $S^2 \times S^1 \times (0,\infty),$
this bundle is a continuous-trace algebra with Dixmier--Douady
invariant $\delta^{\#}.$ Over $\pi^{-1}(0),$ the fiber is
$C^{\ast}(\KR)\otimes \Cpct$ viewed as a $C^{\ast}-$bundle over
$S^1$. The algebra is given by an extension of the form
\begin{equation}\label{SES2}
0 \longrightarrow \hbox{CT}(S^2\times S^1 \times (0,\infty),\delta^{\#})
\longrightarrow
{\mathcal{A}}\rtimes \KR \longrightarrow {\mathcal C}(S^1) \longrightarrow 0,
\end{equation}
where ${\mathcal C}(S^1)$ is $C_0(\KR, \Cpct)$ viewed as a
$C^{\ast}$-bundle over $S^1$ via the quotient map $\KR \to S^1$.
This example can be viewed as calculating the spectrum of the
group $C^{\ast}$-algebra of the motion group $\KC^2 \rtimes
\KR.$

\enlargethispage{2pt}

If we view the crossed product as a $C^{\ast}$-bundle over the
maximal Hausdorff quotient of its spectrum, we obtain a
$C^{\ast}$-bundle over $CS^2 \times S^1.$ Note that $CS^2 \times S^1$
is the physical T-dual. Thus, we define the physical spacetime to be
the maximal Hausdorff regularization of the spectrum of the crossed
product if this spectrum is non-Hausdorff.

Most physical examples of T-duality are built up from the
T-duality of a NS5-brane with a KK-monopole.

Following Ref. \cite{MRCMP}, a topological space $X$ with
$H$-flux $\delta$ may be naturally associated to the
continuous-trace algebra CT$(X,\delta)$. This $H$-flux is
\textit{sourceless}, i.e., we can pick a 3-form which
represents this $H$-flux in a neighborhood of every point of
$X$. However, if the space possesses a \textit{source} of
$H$-flux, we cannot pick such a 3-form in any neighborhood
of the source. (It might be helpful to keep in mind the
description of the Dirac monopole in electromagnetism: recall
that the flux is not defined at the location of the monopole.)
Here, the flux is only defined on $X-Y$ and so we only have a
cohomology class $\delta \in H^3(X-Y,\KZ).$ If we could find a
natural definition of a $C^{\ast}$-algebra $\mathcal{A}$ which
encodes the structure of a space with a source of $H$-flux, then
we hope that the spectrum of its crossed product with $\KR^n$
would still give the T-dual. Note also that in all the physical
examples above, we are dualizing spaces $X$ with a  non-free
$S^1$ action (with $X/S^1 \simeq B$) and no $H$-flux, to spaces
which are trivial $S^1$-bundles over $B$ but contain a source of
$H$-flux.

Assume that we are in the set-up of Section~\ref{SecKKTdual}, with
a space $B \times S^1$ with a source of $H$-flux represented by
a cohomology class $\delta^{\#}$ in $H^3((B-F) \times S^1,\KZ).$
We assume that the source is located somewhere over $F$ on $F
\times S^1 \subset B \times S^1.$ We will replace
continuous-trace algebra CT$(B \times S^1,\delta^{\#})$ by
another $C^{\ast}$-algebra $\mathcal{B}$ so that the maximal
Hausdorff regularisation of $\hat{{\mathcal{B}}}$ is $B \times
S^1$ and we can naturally obtain a $H$-flux in $(B-F) \times
S^1.$ We suppose that $\mathcal{B}$ is an extension of the form
\begin{equation}\label{SES1}
0 \longrightarrow \hbox{CT}((B-F)\times S^1,\delta^{\#}) \longrightarrow {\mathcal{B}} \longrightarrow
{\mathcal{C}}(F) \longrightarrow 0,
\end{equation}
where ${\mathcal{C}}(F) \simeq C_0(\KR) \otimes C_0(F) \otimes
\Cpct.$ (Here, $\mathcal{K}$ is the $C^{\ast}$-algebra of
compact operators on a separable infinite dimensional Hilbert
space.)

For most spaces, there are many extensions $\mathcal{B}$ in
equation~(\ref{SES1}) with the regularisation of ${\hat{\mathcal B}}$
equal to $X$ \cite{JMRExt}. However, in this case, we can see
that $\mathcal{B}$ is uniquely determined.
\begin{lemma}\label{TDualAlg}
Suppose we are given the following data
\begin{itemize}
\item topological spaces $B, F$ such that $F \subseteq B,$
\item a cohomology class  $\delta^\# \in H^3((B-F)\times S^1,\KZ).$
\end{itemize}
Then, topological T-duality for principal $S^1$-bundles with
$H$-flux uniquely determines the algebra $\mathcal{B}$ in
equation~(\ref{SES1}).
\end{lemma}
\begin{proof}
Topological T-duality applied to the trivial principal
$S^1$-bundle\break $(B-F) \times S^1 \to (B-F)$ with $H$-flux
$\delta^{\#}$ gives a principal $S^1$-bundle $E$ over $B-F$. The
characteristic class of this bundle may be obtained by
integrating $\delta^{\#}$ over the $S^1$-fiber.

\enlargethispage{10pt}

This completely determines the T-dual space $X$ up to
equivariant homomorphism as follows. Any orbit of a semi-free
$S^1$-action on a space $X$ can only have two stabilizers
namely, the identity and $S^1.$ As a result, the spaces $X$ can
only have two orbit types: fixed points and free orbits. Hence,
if $F$ is the subset of $B$ such that $\pi^{-1}(F)$ is the fixed
point subset of $X$, by the classification theorem for spaces
with two orbit types (see Ref. \cite{Bredon} Chapter~V Section~5),
the space $X$ is completely specified up to equivariant
homomorphism by $F \subset B$ and the class of the principal
$S^1$-bundle over $B-F.$

Let $\A\,{=}\,C_0(X,\Cpct).$ The $S^1$-action on $X$ may always be
lifted to $\A$ uniquely, and so determines a $C^{\ast}$-dynamical
system $(\A,\alpha)$ up to exterior equivalence. Now, $\mathcal{B}$
may be defined as $\A \underset{\alpha}{\rtimes} \KR.$ The result is
unique up to C$^{\ast}$-algebra isomorphism.
\end{proof}

Note that topological T-duality is a geometric operation on CW
complexes by Ref. \cite{MRCMP,Bunke} and hence may be freely
used in computations. It is clear that $\mathcal{B}$ has a
$\hat{\KR}$-action $\beta$ and the crossed product $\mathcal{B}
\underset{\beta}{\rtimes} \hat{\KR}$ is isomorphic to $\A.$

In general, the spectrum of $\mathcal{B}$ may not be a CW
complex.\footnote{We saw this for the test example above.}We
emphasize that this does not imply that the physical spacetime
is non-Hausdorff, only that for calculational purposes, it is
convenient to take a non-Hausdorff space whose regularization is
the physical spacetime.

Hence, we make the following dictionary
\begin{itemize}
\item Spacetime a principal $S^1$-bundle $X \to B$ with a
sourceless $H$-flux $\delta$ $\Leftrightarrow$ (CT$(X,\delta),
\alpha)$, where $\alpha$ is the lift of the $S^1$-action on $X$
to CT$(X,\delta),$ as in Ref.~\cite{MRCMP}.
\item Space $B\times S^1$ with a NS5-brane of charge $\delta^{\#}$
wrapped on $F \times S^1 \subseteq B \times S^1.$
$\Leftrightarrow$ The unique extension like equation~(\ref{SES1})
above.
\end{itemize}

\enlargethispage{10pt}

If the NS5-brane is not wrapped around a $S^1$-orbit, for
consistency, we should associate to the space an extension like
equation~(\ref{SES1}) above. However, it is not clear how unique such
an extension is. We do not address this question here, since we
are T-dualizing semi-free $S^1$-spaces and the NS5-branes we
encounter will always be wrapped around a $S^1$-orbit.

\section{Dyonic coordinates and KK-monpoles}\label{SecDyon}

\subsection{Physical background}

It might be objected that the result obtained in the previous
section is accidental. To give further evidence that it is
non-trivial, we reproduce the dyonic coordinate of Refs.
\cite{Jens,Harv,Sen} within the current formalism.


In Section~1, we noted that the T-dual of a Kaluza--Klein monopole is
a source of $H$-flux.\footnote{Also termed a $H$-monopole.} If we
apply Buscher's rules \cite{Jens} to a KK-monopole we obtain a
$H$-monopole \textit{smeared} over the $S^1$-fiber over $\{ 0 \}.$
Quantum corrections are expected to localize the $H$-monopole to a
particular point in the fiber. Recall that we identified a
Kaluza--Klein monopole with the space $(\KR^4,g_{\rm TN})$  (termed
``Taub-NUT'' space). This space possesses an isometric action of
$S^1$ with one fixed point and which is $S^1$-equivariantly
\hbox{homeomorphic} to $\KC^2$ with the $S^1$-action $\lambda \cdot
(z_1,z_2) = (\lambda \cdot z_1, \lambda \cdot z_2)$, $\lambda \in S^1
\subseteq \KC, z_i \in \KC$. We have $\KR^4/S^1\simeq \KR^3.$

Suppose we have a KK-monopole located somewhere in $\KR^4.$ By
the\break previous section, this space is a semi-free $S^1$-space. As
argued in the proof of Lemma~\ref{TDualAlg}, a semi-free $S^1$-space $X$ is completely specified by the fixed point set $F
\subset B = X/S^1$ and the class of the principal $S^1$-bundle
over $B-F.$ Thus, it is specified by a class $\lambda$ in
$H^2(B-F).$ \label{SemiFree}

Here, it would be specified by the class of the principal
$S^1$-bundle\break $\KR^4-{\mathbf x} \to \KR^3-{\mathbf x}$, and the
fixed point set ${\mathbf x} \subseteq \KR^3.$ For the case of a
single KK-monopole, the isomorphism class of the principal
$S^1$-bundle is fixed.\footnote{It must be the unique bundle over
$S^2 \times (0,\infty)$ with Chern class $1$.}Thus, we only
need to know ${\mathbf x}$ to fix the KK-monopole space. Hence
only three numbers are needed to specify the
KK-monopole,\footnote{Following Ref.~\cite[page 2--3]{Jens}.}namely the location of the image of the center in $\KR^3$ under
the quotient map. In the T-dual picture, we have a source of
$H$-flux somewhere in the fiber over $\{0\} \times S^1$. This is
specified by the position of the source in $\KR^3$ and the
location of the source in $\{0\} \times S^1.$ Thus four
parameters are needed to specify the T-dual. Since T-dual spaces
are physically equivalent, we should need the same number of
parameters on both sides. It is interesting, therefore, to ask
which datum of the KK-monopole changes when we change the
location of the source of $H$-flux in the $S^1$-fiber of the
T-dual. According to a result of Sen~\cite{Sen} (see also
\cite{Jens,Harv}), this may be obtained as follows. On the total
space of the KK-monopole, we have a zero $H$-flux. This implies
that the gauge field $B$ of the $H$-flux is a closed 2-form
(as $H=dB$). Since $\KR^4$ is contractible, $B$ is also exact.
However, for the multi-Taub-NUT metrics, there
exist\footnote{Note that the total space is noncompact, so the
usual Hodge theorem does not apply here. See Ref. \cite[Section~7]{Haus}.}$L^2$-normalized 2-forms $\Omega_i$ such that
$\Omega_i \not \to 0$ as $|\mathbf{x}| \to \infty.$ It is scalar
multiples of these forms that give rise to non-trivial
$B$-fields. It is this $B$-field that corresponds to the
position of the $H$-monopole in the T-dual. It is termed a
``dyonic coordinate'' in Refs. \cite{Jens,Harv} by analogy with
the case of monopoles in gauge theories \cite{MonopH}. If the
$B$-field changes in time according to $B(t)=\beta(t) \Omega$
then on the T-dual side, (see below) this corresponds to
changing the angular coordinate of the $S^1$ factor of $\KR^3
\times S^1$ via an isometry
\begin{equation}
\kappa(t) = \kappa(0) - \beta(t). \label{TDualB}
\end{equation}

We may explicitly calculate\footnote{We follow Ref.~\cite{Jens}
here.}the above effect using equation~(\ref{EqBuscher}). Recall, that
the Taub-NUT metric was
\begin{align*}
g_{\rm TN}(r,\theta,\phi,\kappa)&=H d\vec{r}\cdot d\vec{r} +
H^{-1}(d\kappa^2+ (1-\cos(\theta))d\phi d\kappa\\
&\quad+ (1-\cos(\theta))^2 d\phi^2).
\end{align*}
Its T-dual in the absence of a $B$-field was the $H$-monopole
metric
\[
g_{H}(r,\theta,\phi,\kappa)=H(d\kappa)^2+H(dr)^2+Hr^2\sin^2(\theta)(d\phi)^2+
Hr^2(d\theta)^2
\]
Note that for the Taub-NUT metric, $g_{00}=H^{-1};$ the harmonic
form $B$ discussed above is given by
\begin{align}\label{EqOmega}
B & = \beta \Omega= \beta d\Lambda = \beta d \left(\frac{1}{g^2H}\left(d\kappa
+ \frac{(1-\cos(\theta))}{2} d\phi\right) \right) \nonumber \\
& = -\frac{\beta H'}{g^2 H^2} dr \left(d\kappa +
\frac{(1-\cos(\theta))}{2} d\phi\right)
- \frac{\beta}{g^2 H} \sin\,{\theta} d\theta\,d\phi
\end{align}
Hence we have $b_{01}= -\frac{(\beta H')}{(g^2 H^2)}$ and so
\[
\tilde{g}_{00}=H, \tilde{g}_{01}= - \frac{(\beta H H')}{(g^2 H^2)}= -
\frac{(\beta H')}{(g^2 H)}.
\]
Thus, the T-dual is given by
\begin{align*}
\tilde{g} &= H(r) \left(d\kappa^2- \frac{(\beta H')}{(g^2 H)} d\kappa\, dr\right) + \{
\hbox{terms of } g_{\alpha \beta}, \beta \neq 0, \alpha \neq 0 \} \\
& = H(r)(d\kappa^2 - (\beta H')(g^2 H^2)d\kappa \, dr +
(\beta^2 {H'}^2)/(4 g^4 H^4) dr^2) + \cdots \\
& = H(r)( (d\kappa - (\beta H')/(g^2 H^2) dr)^2) + \cdots \\
& = H(r)(d(\kappa + \beta/(g^2 H)))^2 + \cdots
\end{align*}

Therefore, if we take a diffeomorphism $\Gamma$ of the T-dual,
$\Gamma: \KR^3 \times S^1 \to \KR^3 \times S^1$ given by
\[
r = r, \theta = \theta, \phi = \phi, K = \kappa +
\frac{\beta}{(g^2H(r))},
\]
we see that $\tilde{g}(r, \theta, \phi, \kappa)=
\Gamma^{\ast}(g_{H}(r,\theta,\phi, K)).$ Also, $\Gamma$ is an
isometric diffeomorphism\footnote{Therefore, by the principle
of general covariance, these two are indistinguishable at the
level of general relativity, as expected.}between the distinct
Riemannian manifolds $(\KR^3 \times S^1, g_{H})$ and $(\KR^3
\times S^1, \tilde{g})$. Note that as $r\to \infty,$ $\Gamma$
approaches the isometry $\kappa \to \kappa + \beta.$ In general,
it is preferable if such transformations approach the identity
at infinity. Thus, we consider instead the diffeomorphism
$\Lambda$ of the T-dual, $\Lambda: \KR^3 \times S^1 \to \KR^3
\times S^1$ given by
\[
r=r,\theta=\theta,\phi=\phi,\tilde{K} = \kappa +
\frac{\beta}{(g^2H(r))}- \beta,
\]
which approaches the identity as $r \to \infty$, we obtain that
\[
\tilde{g}(r, \theta, \phi, \kappa) =
\Lambda^{\ast}(g_{H}(r,\theta,\phi, \tilde{K} + \beta)).
\]

We expect that it should be possible to model the dyonic coordinate
discussed above within the formalism of Refs. \cite{MRCMP,BouEvMa}.
We expect this because the above T-dual was also obtained from
Buscher's rules. If we could mimic this effect within the topological
formalism, this would give added evidence that the T-dual in
Chapter~(1) is the ``correct'' one.\vspace*{-6pt}

\subsection{Mathematical formalism}

We first simplify the problem by passing to a suitable
compactification. We view $\KC^2$ as an open subset of $\KCP^2:$
i.e., we have compactified $\KC^2$ by adding an $S^2$ at
$\infty$ obtained by collapsing each $S^1$-orbit to a point. As
$H^2(\KCP^2,\KZ) \simeq \KZ,$ and $\KCP^2$ is compact, there is
a unique harmonic form on $\KCP^2$ whose integer multiples
generate $H^2(\KCP^2,\KZ) \subseteq H^2(\KCP^2, \KR).$ It is
shown in Ref. \cite{Haus} that the restriction of this form to
$\KC^2 \subseteq \KCP^2$ is related to $\Omega.$ Also $\beta(t)
\Omega$ extends to a closed distributional 2-form on $\KCP^2$
and gives rise\ \cite{Haus} to an element of $H^2(\KCP^2, \KR).$
It is not clear which topological object may be associated with
a real cohomology class. However, if we restrict ourselves to
integral $B$-fields (i.e., elements of the form $m \Omega, m \in
\KZ$) we may reformulate the above as follows. If we change the
$B$-field by adding an element of $H^2(\KCP^2, \KZ)$ to it, on
the T-dual side, this should correspond to rotating the
$S^1$-fiber via equation~(\ref{TDualB}).

We use homogenous coordinates $[ x_1{:}x_2{:}x_3 ]$ on $\KCP^2$ (with
$x_i \in \KC$ and\break $( x_1{:}x_2{:}x_3) \sim (\lambda x_1{:}\lambda x_2{:}
\lambda x_3), \forall \lambda \in \KC^{\ast}$) Then $\KC^2$
corresponds to the subset
\[
U = \left\{ [x_1{:}x_2{:}x_3] \mid x_3 \neq 0 \right\}
\]
and the sphere at infinity to the subset
\[
W = \left\{ [x_1{:}x_2{:}x_3] \mid x_3 = 0 \right\}.
\]
We consider the action $\lambda \cdot [x_1{:}x_2{:}x_3] = [\lambda
x_1{:}\lambda x_2 {:}x_3]$, $\lambda \in S^1$ on $\KCP^2.$ This is
obviously well defined. Note that on $U$ it turns into the action of
Section~2 because the following commutes
\begin{equation}\label{CDCP2}
\CD
[x_1{:}x_2{:}x_3]@>>> [\lambda x_1{:}\lambda x_2{:} x_3]\\
@VVV  @VVV \\
(x_1/x_3,x_2/x_3) @>>> (\lambda x_1/x_3, \lambda x_2/x_3).
\endCD
\end{equation}
Hence we have an $S^1$-action on $\KCP^2$ which we may lift to a
$\KR$-action $\alpha_t$ on $C(\KCP^2,\Cpct).$ Any 2-form $\lambda \in
H^2(\KCP^2, \KZ)$ gives rise to a spectrum-fixing automorphism of
$C(\KCP^2,\Cpct).$ We recall that spectrum-fixing automorphisms of
$C^\ast-$algebras are classified up to inner automorphisms by their
Phillips-Raeburn invariant which is a homomorphism
$\zeta:\hbox{Aut}_{C_0(X)}(\A) \to H^2(X,\KZ).$ The following theorem
shows that this automorphism may always be ``dualized'':

\begin{theorem}\label{ThmDyonCP2}
Let $X$ be a finite CW complex with a semi-free $S^1$-action. Let
$\alpha_t$ be a lift of the $S^1-action$ on $X$ to $C(X,\Cpct).$

\begin{enumerate}
\item[1.] Let $[\lambda ] \in H^2(X,\KZ).$ Then, there is a
$\KR-$action $\beta$ on $C(X,\Cpct)$ exterior equivalent to $\alpha$
and a spectrum-fixing $\KZ$-action $\lambda$ on $C(X,\Cpct)$ which
has Philips--Raeburn obstruction $[ \lambda ]$ such that $\beta$ and
$\lambda$ commute.
\item[2.] With the notation above, the action $\lambda$ induces a
$\KZ$-action $\tilde{\lambda}$ on $C(X,\Cpct)\break
\underset{\alpha}{\rtimes} \KR.$ The induced action on the crossed
product is locally unitary on the spectrum of the crossed product and
is thus spectrum fixing.
\end{enumerate}
\end{theorem}

\begin{proof}
Let $A = C(X,\Cpct)$ then $X = \hat A.$
\begin{enumerate}
\item[1.]
We have a short exact sequence
\[
0 \longrightarrow  \hbox{Inn}(A) \longrightarrow  \hbox{Aut}_{C_0(X)}(A) \overset{\zeta}{\longrightarrow } H^2(X,\KZ)
\longrightarrow  0.
\]
Pick any $\KZ$-action$ \tilde{\lambda}$ with $\zeta(\tilde{\lambda})=
[\lambda]$. Note that $\zeta(\alpha_{-t} \tilde{\lambda} \alpha_{t})
= \alpha_t^{\ast}(\zeta(\lambda))$ (By Lemma~4.4 of
Ref.~\cite{CKRW}). Since $\alpha_t^{\ast} = id, \forall t \in \KR$
(as $\alpha_{1}^{\ast} = id$ ), we have $\zeta(\alpha_{-t}
\tilde{\lambda} \alpha_{t}) = [\lambda], \forall t \in \KR$. We see
that $\alpha_{-t} \lambda \alpha_t$ is equal to $\lambda$ up to inner
automorphisms. If it were exactly equal to $\lambda$ we would obtain
a map $\phi:\KR \times \KZ \to \hbox{Aut}(A).$ As it is, we obtain a
map ${\tilde \phi}:\KR \times \KZ \to \hbox{Out}(A).$ To lift this to
Aut$(A)$, by Ref.~\cite{CKRW}, Lemma $4.6$, an obstruction class in
$H^3_M(\KR \times \KZ, C(X, \KT))$ must vanish. To calculate this
cohomology group, we use the fact that $H^3_M(G,A) \simeq
\underline{H}^3_M(G,A)$ if $G$ is second countable and locally
compact and $A$ is a polish abelian $G$-module
(see~\cite[Theorem~7.4]{RaeWil}). Now, by
Ref.~\cite[Theorem~9]{Moore3}, there is a spectral
sequence\footnote{See also Ref.~\cite[page~190]{RaeWil}.}converging
to ${\underline H}^*(\KR \times \KZ,C(X,\KT))$ whose $E^2$ term is
$E^2_{p,q}={\underline H}^p(\KZ,{\underline H}^q(\KR,C(X,\KT))$. By
Ref.~\cite[Theorem~4.1]{RaeRos}, we have that
$H^q(\KR,C(X,\KT))=0,q>1$. Note that the $\KZ$-module
$H^{\ast}_M(\KR, C(X,\KT))$ has a trivial $\KZ$-action, since
$C(X,\KT)$ has a trivial $\KZ$-action. Since $\KZ$ has the discrete
topology, the Borel cochains with values in a Polish abelian group
$A$ with a trivial $\KZ$-action are simply all the cochains.
Therefore, $H^{\ast}_M(\KZ,A) \simeq H^{\ast}(\KZ,A),$ where the last
cohomology group is calculated by the usual group cohomology theory.
Since $H^{k}(\KZ,A) = 0$ for $k > 1,$ we see that $E^{p,q}_2 = 0$ for
$p > 1$ as well as for $q > 1.$ Thus $E^{p,q}_2 = 0$ for all $p+q =
3$ and $H^3$ vanishes. As a result, the action $\tilde{\phi}$ lifts to a
twisted action $\phi'.$ By Raeburn's Stabilization Trick, this is
exterior equivalent to an ordinary action $\phi.$


Note that the restriction of $\phi$ to the $\KR$-factor gives an
$\KR$-action exterior equivalent to $\alpha$ (since the lift of
the $S^1$-action is unique up to exterior equivalence).

\item[2.] Note that $\lambda$ is a locally unitary action. Hence, we
may pick a sufficiently small open set $U \subset X$ such that
$\lambda$ is unitary on the localization $\A_U$ of $\A$ to $U.$ This
defines an element $f \in C_b(\KR,\hbox{UM}(\A))$ by $f(t) =
u_{\alpha}, \forall t$. Note that $U$ may be taken to be
$S^1-$invariant. Since there are two orbit types and we're assuming
everything is homotopy finite, we can choose $U$ to be equivariantly
homeomorphic to either $S^1 \times V$ with $V$ contractible (if we're
away from a fixed point) or to a cone times $V,$ $V$ a contractible
open subset of the fixed set. For both these, $H^2(U,\KZ) \simeq 0,$
so $\lambda$ localized to these sets is unitary. Now, since $U$ is a
union of $S^1$-orbits, $C_b(\KR,\hbox{UM}(\A)) \subseteq \hbox{UM}(\A
\underset{\lambda}{\rtimes} \KR)$. The induced automorphism on $\A
\underset{\lambda}{\rtimes} \KR$ is given on $C_0(\KR,\A)$ (which is a
dense subspace of $\A \underset{\lambda}{\rtimes} \KR$) by
$\tilde{\lambda}(g)(s) = f(s)^{\ast}g(s) f(s), \forall s \in \KR,$
this extends to a unitary automorphism of $\CrPr{\A_U}{\KR}{\alpha}.$
Thus $\tilde{\lambda}$ is locally unitary on the spectrum of the
crossed product.
\end{enumerate}
\end{proof}

Thus we see that under T-duality, a class in $H^2(\KCP^2,\KZ)$
gives rise to a spectrum fixing automorphism of the crossed
product algebra. We identify this with a rotation of the form
equation~(\ref{TDualB}) with $\beta = 2m\pi, m \in \KZ.$

Note that in our example, the spectrum $X^{\#}$ of the crossed
product is not Hausdorff. Hence, the crossed product algebra is not
continuous trace. Thus, this spectrum fixing automorphism does not
define a cohomology class in $H^2(Y,\KZ)$ where $Y$ is the Hausdorff
regularization of $X^{\#}.$ Physically, this is reasonable, since
there is an $H$-flux in the T-dual so we would not expect a $B$-field
there.

\subsection{Multiple KK-monopoles}

Consider the Multi-Taub-NUT space\footnote{defined in
Chapter~1.}$X_k$. Using the coordinates given there,\footnote{see
equation~(\ref{EqMultKK}).}the metric on $X_k$ is given by
\begin{gather*}
g_{k{\rm TN}} = H(\vec{\mathbf r}) d\vec{\mathbf r} \cdot d\vec{\mathbf r} +
H(\vec{\mathbf r})^{-1} \left(d\kappa + \frac{1}{2} \omega\cdot
d\vec{\mathbf r} \right)^2 \nonumber \\
\hbox{where } H(\vec{\mathbf r}) = 1 + \sum^{k}_{i=1}
\frac{1}{|\vec{\mathbf r}-\vec{{\mathbf r}_i}|}.
\end{gather*}
The T-dual is given by
\begin{equation}
g_{H_k}(r,\theta,\phi,\kappa) = H ( (d\kappa)^2 + d\vec{r}\cdot
d\vec{r})
\end{equation}
The harmonic forms $B_k$ on $X_k$ are given by
\begin{gather}
B_i = \beta \Omega_i = \beta d\xi_i, \nonumber \\[-1pt]
\xi_i = \alpha_i - \frac{H_i}{H} (d\kappa + \alpha), \nonumber \\[-1pt]
H_i = \frac{1}{|r - r_i|}, \nonumber \\[-1pt]
B_i = -\beta \frac{\partial}{\partial
r}\left(\frac{H_i}{H}\right)
 d\kappa\,dr + \left\{ \protect{\hbox{terms not containing }}d\kappa \right\} ,
\nonumber \\[-1pt]
B_{01i} = -\beta \frac{\partial}{\partial r}\left( \frac{H_i}{H}
\right)
\end{gather}


\noindent Hence, for the T-dual metric,
\[
g_{00} = H,\quad g_{01} = -\beta \frac{\partial}{\partial r}\left(
\frac{H_i}{H} \right)
\]
As in the previous section, the T-dual is given by
\[
g_{H_k} = H(r) \left(d\left(\kappa + \beta \frac{\partial}{\partial
r}\left( \frac{H_i}{H} \right)\right)\right)^2 + \left\{
\protect{\hbox{terms of }} g_{\alpha \beta},\quad \alpha \neq
0,\enspace \beta \neq 0 \right\}.
\]
We prefer to take the following as a basis for the set of harmonic
forms
\[
\tilde{B_i} = \sum_{j, j\neq i} B_i = - \beta \frac{\partial
}{\partial r} \left( 1- \frac{H_i}{H} \right) d\kappa dr + \cdots
\]
Then, the T-dual is given by
\[
g_{H_k} = H(r) \left(d\left(\kappa + \beta \frac{\partial}{\partial
r}\left( 1 - \frac{H_i}{H} \right)\right)\right) + \left\{\protect{
\hbox{terms of }} g_{\alpha \beta},\quad \alpha \neq 0,\enspace \beta
\neq 0 \right\}.
\]

We take the diffeomorphism $\Gamma$ of $\KR^3 \times S^1$ given by
\begin{gather}
r = r, \nonumber \\
\theta= \theta, \nonumber \\
\phi = \phi, \nonumber \\
\kappa = \kappa + \beta \frac{\partial}{\partial r} \left( 1 -
\frac{H_i}{H} \right) - \beta. \nonumber
\end{gather}
Hence, exactly in the previous section, $g_{{\rm
TN}_k}(r,\theta,\phi, \kappa) =
\Lambda^{\ast}(g_{H_k}(r,\theta,\break\phi, \kappa + \beta)).$

The calculation in the previous subsection trivially extends to
multi-Taub-NUT spaces. We need a suitable compactification
$\tilde{X_k}$ of $X_k$ such that the harmonic forms on $X_k$ are
related to the cohomology of $\tilde{X_k}.$ Since $g_{{\rm TN}_k}$ is
a metric of fibered boundary type \cite{Haus}, we use the
compactification given in that paper. Thus, $\tilde{X_k}$ is obtained
by collapsing the $S^1$-fibres of $S^3/\KZ_k$ (which is the boundary
of $X_k$ at $\infty$) to points to obtain an $S^2.$ We find from
\cite{Haus} that the $L^2$-harmonic forms on $X_k \subseteq
\tilde{X}_k$ are related the harmonic forms which are the elements of
$H^2(\tilde{X}_k,\KZ).$ Now, $X_k$ has an $S^1$-action which extends
to $\tilde{X_k}$ by fixing the $S^2$ at $\infty.$ In Theorem
(\ref{ThmDyonCP2}), applies to any finite CW complex $X$ which
possesses an $S^1$-action. This would be be the analogue of the
dyonic coordinate for $X_k.$ In general, we can repeat the above
construction for any Riemannian manifold of fibered boundary type
using the compactification in Ref.~\cite{Haus}.

\section{Gerbes and T-duality \label{SecGerbe}}

In this section, we show that we may naturally associate a $2$-gerbe
(see below) to a semi-free $S^1$-space. Similarly, we show that we
may associate a $3$-gerbe to a space with a source of $H$-flux. We
show that topological T-duality induces a natural map between these
two gerbes.


We are trying to model a space containing a source of $H$-flux, a
$3$-form field. It is useful to begin by studying a simpler example,
a Dirac monopole (this is a source of a $2$-form field, the
electromagnetic flux). We begin by reviewing a construction of J.-L.
Brylinski \cite{Bryl1}.

\subsection{The Dirac monopole}

It is well known that a Dirac monopole situated at ${\rm x} \in
\KR^3$ defines a line bundle $E$ on $\KR^3-{\rm x},$ the gauge
bundle, together with a connection $\nabla_{X}$ on this line bundle.
The connection is only specified up to a gauge transformation, i.e.,
we consider $\nabla_{X}$ equivalent to $U\nabla_{X} U^{\ast}$, where
$U$ is a section of the bundle $\operatorname{End}(E).$

The curvature of this connection is identified with the
electromagnetic field strength $F$ emitted by the monopole. It is a
closed 2-form on $\KR^3-{\rm x}.$ Following Refs.
\cite{Bryl1,Witten1}, we view the field strength of the monopole $F$
as a distribution-valued $2$-form on $\KR^3$ with singular support at
${\rm x}.$ Maxwell's equations for the field generated by the
monopole are $dF = l$ and $d\ast F = h$, where $l,h$ are
distribution-valued differential forms  with support at the point
${\rm x}$ in $\KR^3;$ these model the monopole as a source of the
electromagnetic field.

(From now on, we assume that the monopole is located at ${\rm x} =
0.$ We will relax this restriction later.) From the equations
governing $F$, it follows that $F$ diverges as $F \sim 1/|{\rm
y}|^{\alpha} $ for some constant $\alpha$ as ${\mathbf y} \to 0$

Classically, a line bundle with connection has a characteristic class
which is the de Rham cohomology class of its curvature 2-form. In
Refs. \cite{Witten1,Bryl1}, the authors obtain a generalization of
this characteristic class for the monopole as follows. Suppose we
attempt to use the curvature of the line bundle associated to the
monopole to obtain a cohomology class. This class would most
naturally reside in the cohomology of the complex
$\Omega_{\{0\}}^{\ast}(\KR^3)$ which consists of distribution-valued
differential forms on $\KR^3$ which have singular support at $0 \in
\KR^3.$ It can be shown that the cohomology of this complex is
$H^{\ast}(\KR^3,\KR^3-{0}).$ Using the long exact sequence of the
pair $(\KR^3,\KR^3-{0}),$ we obtain the exact sequence
\begin{equation}
H^2(\KR^3)\longrightarrow H^2(\KR^3-{0})\longrightarrow
H^3(\KR^3,\KR^3-{0})\longrightarrow H^3(\KR^3).
\end{equation}\vskip6pt
Since all the cohomology groups of $\KR^3$ vanish, we obtain an
isomorphism $H^2(\KR^3-{0}) \to H^3(\KR^3, \KR^3-{0})$ under which
$[F|({\KR^3-\{0\}})] \to [dF=l].$ Now, as explained in Ref.
\cite{Bryl1} we would like to move the monopole about $\KR^3$ without
affecting the above class. This may be done by passing to $S^3$
(which is to be viewed as) the 1-point compactification of $\KR^3.$
We assume given an inclusion $i:\KR^3\to S^3$ which induces maps
$i^{\ast},j^{\ast}$ as shown below.
{\fontsize{10}{11}\selectfont\begin{equation} \CD
H^2(\KR^3) @>>> H^2(\KR^3-{0}) @>>> H^3(\KR^3,\KR^3-{0})@>>>H^3(\KR^3)@. \\
@AAi^{\ast}A    @AAj^{\ast}A       @AAk^{\ast}A         @AAi^{\ast}A  @.\\
H^2(S^3)   @>>> H^2(S^3-{0})   @>>> H^3(S^3,S^3-{0}) @>>>H^3(S^3)
@>>> 0
\endCD
\label{CDia1}
\end{equation}}
Here $k^{\ast}$ is an isomorphism by excision. This gives the
following commutative diagram with exact rows
\begin{equation}
\CD
0 @>>> \KZ @>>> \KZ @>>> 0 \\
@AAi^{\ast}A @AAj^{\ast}A @AAk^{\ast}A  @AAi^{\ast}A \\
0 @>>> 0 @>>> \KZ @>>> \KZ
\endCD
\label{CDia1a}
\end{equation}
and so an isomorphism of $H^3(\KR^3,\KR^3-{0})$ with $H^3(S^3).$


Hence, given a monopole situated at ${0}$ in $\KR^3$, we view its
curvature 2-form as a distribution-valued 2-form defined on $\KR^3$
with singular support at ${0}.$ Its cohomology class gives an element
of $H^3(\KR^3,\KR^3-{0})$ which, by the above isomorphism, gives a
class in $H^3(S^3).$ It is also clear by the above argument that
changing the location of the monopole from $0$ to ${\rm x} \in \KR^3$
will not change the above class in $H^3(S^3)$.

It is interesting to view the class obtained above in a geometric way
which will be useful later. Elements of $H^3(S^3)$ are in one-to-one
correspondence with stable continuous-trace $C^{\ast}$-algebras with
spectrum $S^3$. Given a monopole on $\KR^3$, is it possible to
uniquely obtain a continuous-trace algebra on $S^3$? Also, does every
continuous-trace algebra on $S^3$ arise in this way?

Suppose we are given a monopole located at a point ${\rm x} \in
\KR^3$. This gives rise to a closed 2-form (the gauge field strength)
on $\KR^3-{\rm x}$. The de Rham cohomology class $\omega$ of this
2-form defines an element\footnote{The form is integral because of
``Dirac Quantization''.} in $H^2(\KR^3-{\rm x}, \KZ)$ and thus a map
$(\KR^3-{\mathbf x}) \to K(\KZ,2).$ Since $K(\KZ,2)$ is homotopy
equivalent to $\mathcal{PU}$ we obtain a map from $(\KR^3-{\rm x})
\to \mathcal{PU}.$ (Note that $\KR^3-{\rm x} \simeq S^3 - \{ {\rm x},
\infty \}$.) Using the gluing construction of continuous-trace
algebras on $S^3$ given in Ref. \cite{JMRExt}, we see that we obtain
a unique stable continuous-trace algebra on $S^3$ associated to the
monopole.\footnote{Take trivial continuous-trace algebras over
$S^3-x$ and $S^3-{\infty}$ and glue on the overlap using the above
function, noting that ${\mathcal PU} \simeq \hbox{Aut}(\Cpct)$.}
Every class in $H^3(S^3)$ arises in this way because the
Dixmier-Douady invariant of the continuous-trace algebra, which
classifies the algebra up to isomorphism, is the image of $\omega$ in
$H^3(S^3)$ via the isomorphism $H^2(S^2)\to H^3(S^3)$.

\looseness=-1 Note that the same algebra on $S^3$ represents the
gauge bundle of a\break monopole at any other\footnote{Hence the
associated continuous-trace algebra does not determine the monopole
completely.} point $x$ in $\KR^3$. Pick as the open sets $U_{\rm x} =
S^3-{x}$ and $U_{\infty} = S^3 - {\infty}$. Since these open sets are
contractible, the algebra localized to these sets is
$C_0(U_{\ast},\Cpct).$ We obtain transition functions on $U_{\mathbf
x} \cap U_{\infty}$ which define the same class in $H^2(S^2)$ as
above, since the Dixmier--Douady invariant is the same. This shows
that we have a line bundle defined on $U_{\rm x} \cap U_{\infty}$
which we identify with the gauge bundle of a monopole
located~at~${\rm x}$.\vspace*{-4pt}

\subsection{Sources of $H$-flux}

For the first T-dual pair, we have a source of $H$-flux situated at
${0}\times S^1$ on $C^{0}S^2 \times S^1.$ We represent such a
situation by a distribution-valued 3-form on $C^0S^2 \times S^1$ with
singular support at ${0}\times S^1.$ As above, we expect that its
cohomology class in a suitably defined group should give a
topological invariant of this situation. By an argument similar to
the one given above, the cohomology class should lie in $H^4(C^0S^2
\times S^1, {0} \times S^1)$. We have an exact sequence,\vspace*{3pt}
\begin{equation}
\minCDarrowwidth1pc \CD
H^3(C^0S^2\times S^1)@>>>H^3((C^0S^2-{0})\times S^1) \\
@. @V\phi^{\ast}VV \\
@. H^4(C^0S^2 \times S^1,(C^0S^2-{0})\times S^1) @>>>
H^4(C^0S^2\times S^1)
\endCD
\label{LESPhi}
\end{equation}

As $C^0S^2 \times S^1$ is homotopy equivalent to $S^1$, we get an
isomorphism $\phi^{\ast}:H^3((C^0S^2-{0})\times S^1) \to H^4(C^0S^2
\times S^1, (C^0S^2-{0})\times S^1))$. Now consider the inclusion
$C^0S^2 \times S^1 \to S^3 \times S^1$, where we view $S^3$ as the
1-point compactification of $C^0S^2.$ We get a commutative
diagram\vspace*{3pt}
\begin{equation}
\CD H^3(C^0S^2\times S^1) @>>> H^4(C^0S^2\times
S^1,(C^0S^2-{0})\times S^1)
@>>> \\
@AAi^{\ast}A      @AAj^{\ast}A \\
H^3((S^3-{0})\times S^1) @>>> H^4(S^3\times S^1,(S^3-{0})\times S^1) @>>> \\
H^4(C^0S^2\times S^1)@>>>H^4((C^0S^2-{0})\times S^1) \\
@AAk^{\ast}A  @AAi^{\ast}A \\
H^4(S^3 \times S^1) @>>> H^4((S^3-{0})\times S^1)
\endCD
\label{CDia2}
\end{equation}
and hence an isomorphism $H^4(C^0S^2\times S^1,(C^0S^2-{0})\times
S^1)\,{\simeq}\,H^4(S^3\times S^1)$.

We would like a geometric version of this isomorphism; i.e., given an
extension like equation~(\ref{SES2}), we would like to naturally
associate a class in $H^4(S^3\times S^1)$. We use the above argument
to associate a $2$-gerbe (see below) on $B^{+}$ to a semi-free
$S^1$-space $X$ with $X/S^1=B.$ We show in Theorem. (4.3)
below that T-duality gives a natural mapping between $2$-gerbes on
$B$ and $3$-gerbes on $B^{+} \times S^1$. The characteristic class of
this $3$-gerbe is exactly the class obtained above.\vspace*{-5pt}

\subsection{Gerbes}

Following Ref.~\cite{Hitch}, we define a $k$-gerbe on a space to
be a geometric object whose isomorphism classes are naturally
associated to a class in $H^k(X;\underline{\KC^{\ast}}) \simeq
H^{k+1}(X; \KZ)$ where $\underline{\KC}^{\ast}$ is the sheaf of
germs of $\KC^{\ast}-$valued functions on $X.$
Therefore,\footnote{The word ``gerbe" in the following always
refers to gerbes in the sense of Ref. \cite{Hitch} i.e., strict
gerbes in the sense of Brylinski.} a $0$-gerbe is just an element
of $C(X, \KC^{\ast})$ up to homotopy. A $1$-gerbe is a line bundle
(since isomorphism classes of such objects are in one-to-one
correspondence with elements of $H^1(X;\underline{\KC^{\ast}})$).
A\,$2$-gerbe is an object whose isomorphism classes are naturally
associated to an element of $H^2(X;\KC^{\ast}) \simeq H^3(X;\KZ)$
and may be identified with a continuous-trace algebra on $X$ up to
isomorphism. No explicit realizations of gerbes above degree $2$
are known, but they may be easily specified (see Ref.~\cite{Chatt}
Section~4.5) in terms of data similar to that given below for
$2$-gerbes.

\begin{definition}\rm
A abelian, locally-trivialized $2$-gerbe\footnote{We follow
Ref.~\cite{Chatt} Definition~$2.1.1$, here. These are termed ``gerbs"
in Ref.~\cite{Chatt} and are shown to be identical with $2$-gerbes in
the sense of Ref.~\cite{Hitch} later on in that paper.} on a space
$X$ is specified by the following data
\begin{itemize}
\item An open cover of $X$
\[\{U_{i}:i\in I \} \quad \hbox{with}\quad \bigcup_I U_i = X\]
(and we write $U_{i,j}= U_i \cap U_j$ and so on for other tuples
of indices).
\item A complex line-bundle $\Gamma^{i}_{j}$ over
$U_{i,j}$ for each ordered pair $(i,j),i\neq j,$ such that
$\Gamma^{j}_{i}$ and $\Gamma^{i}_{j}$ are dual to each other.
\item For each ordered triple of distinct indices $(i,j,k),$ a
nowhere-zero\break \hbox{section}
$$\theta_{i,j,k} \in \Gamma(U_{i,j,k};\Gamma^{j}_{i}\otimes\Gamma^{k}_{j}
\otimes \Gamma^{i}_{k})$$ such that the sections
$\theta_{i,j,k}$ of reorderings of a triple $(i,j,k)$ are
related in the natural way.
\item On 4-fold intersections we require that $\delta \theta = 1$,
where $\delta$ is the Cech coboundary operator.
\end{itemize}
\label{2GerbDef}
\end{definition}

We refer to Ref.~\cite{Chatt} for the notion of a refinement of a
$2$-gerbe and a proof of the fact that a $2$-gerbe naturally gives
rise to a class in $H^3(X;\KZ)$. Note that by passing to a
sufficiently fine cover, the $\Gamma^i_j$ could be trivialized. Then,
the above definition reduces to that of a Cech 3-cocycle. However,
the above definition works for any open cover.

Note that the above definition of a $2$-gerbe may be used to
construct a continuous-trace algebra on $X$. The vector bundles
$\Gamma^i_j$ give maps from $U_{i,j}$ to ${\mathcal PU}$. Since
${\mathcal PU}$ is isomorphic to $\hbox{Aut}(\Cpct)$, these maps may
be used to glue $C_0(U_i,\Cpct)$ together along the $U_{i,j}$ to get
a continuous-trace algebra as in Ref.~\cite{JMRExt}. Conversely,
given a continuous-trace algebra on $X$, we obtain a gerbe, since the
${\mathcal PU}$ cocyles defining the continuous-trace algebra will
give the vector bundles $\Gamma^i_j$. The remaining conditions are
automatically satisfied, by definition. In particular, the image of
the cohomology class of $\theta_{i,j,k}$  via the Bockstein map will
be the Dixmier--Douady invariant.

\begin{definition}\rm
A locally trivialized $3$-gerbe\footnote{We follow Ref.~\cite{Chatt}
Section~$4.5$ here.} on a space $X$ consists of the following data:
\begin{itemize}
\item An open cover of $X$
\[\{U_{i}:i\in I \} \quad \hbox{with}\quad \bigcup_I U_i = X\]
(and we write $U_{ij}= U_i \cap U_j$ and so on for other tuples
of indices).
\item A $2$-gerbe, i.e., a continuous-trace algebra ${\mathcal A}^{j}_{i}$ over
$U_{ij}$ for each ordered pair $(i,j),i\neq j$, such that the classes
of ${\mathcal A}^{j}_{i}$ and ${\mathcal A}^{i}_{j}$ in $H^3(U_{ij},
\KZ)$ are inverses of each other.
\item A canonical trivialization $\Gamma_{ijk}$ of the tensor product
${\mathcal A}^{j}_{i}\vert_{U_{ijk}} \otimes {\mathcal
A}^{k}_{j} \vert_{U_{ijk}} \otimes {\mathcal A}^{i}_{k}
\vert_{U_{ijk}}$ (This would be a line bundle.) The line bundles
$\Gamma$ are related in the natural way under reorderings of
$(i,j,k).$
\item A trivialization of the coboundary of the $\Gamma_{ijk}$ on
4-fold intersections $U_{ijkl},$ i.e., a canonical section
$\eta_{ijkl}$ of
\[\Gamma_{ijk}\vert_{U_{ijkl}}\otimes \Gamma_{ijl}^{-1}\vert_{U_{ijkl}}
\otimes \Gamma_{ikl}\vert_{U_{ijkl}} \otimes
\Gamma_{jkl}^{-1}\vert_{U_{ijkl}}\] and all the sections $\eta$ are
related in the natural way under reorderings of $(i,j,k,l)$
\item On 5-fold intersections, we require that $\delta \eta = 1$,
where $\delta$ is the Cech coboundary operator.
\end{itemize}
\label{3GerbDef}
\end{definition}

The characteristic class of this $3$-gerbe is the cohomology class of
$\eta \in H^4(X;\KZ)$. (Thus a $3$-gerbe would define a cohomology
class in $H^4(X; \KZ)$ in exactly a similar manner as a $2$-gerbe
determines a cohomology class in $H^3(X;\KZ)$.) From the above
definitions, we can see that a $k$-gerbe possesses $(k-1)$-gerbes as
``local sections''. For example, a non-trivial line bundle (a
$1$-gerbe) has continuous functions as local non-zero sections.
Similarly, a continuous-trace algebra (a $2$-gerbe) has local {\it
objects} which are line bundles. This may be seen as follows.
Continuous-trace algebras satisfy Fell's condition, which guarantees
the existence of a local rank-one projection in some neighborhood
$U_x$ of each $x \in X$. This is the same as a map $U_x \to
\hbox{Gr}(1,{\mathcal H}_x)$ for some Hilbert space ${\mathcal H}_x.$
However, $\hbox{Gr}(1,{\mathcal H}_x)$ is the classifying space for
line bundles over $U_x.$ The algebra is trivial if and only if there
is a global rank-one projection. This would be the same as specifying
a global line bundle on $X$ which would be the analogue, for a
$2$-gerbe, of a global section of a line bundle. Similarly, a
$3$-gerbe on $X$ (which would be classified by an element of
$H^4(X;\KZ)$) would have stable continuous-trace algebras as local
sections. In the case of a line bundle, it is impossible to pick a
global non-zero section unless the bundle is trivial, similarly, for
a gerbe it is impossible to pick a global object, unless the gerbe is
trivial. Hence, $2$-gerbes may be used to study situations in which
we have ``partially defined" line bundles. This is the case, for
example, in the monopole of the previous section. It also explains
why we could naturally associate to it a class in $H^3.$ Similarly,
we expect $3$-gerbes to be useful for describing situations where we
have ``partially defined" continuous-trace algebras that is, sources
of $H$-flux.

\subsection{Application to T-duality}

We consider a special case of T-duality formulated in
Ref.~\cite{MRCMP}, Lemma $4.5$. Continuous-trace algebras on $B\times
S^1$ are T-dual to $U(1)$-bundles $E$ on $B$. We may restate this by
saying that T-duality gives a correspondence between $1$-gerbes on
$B$ and $2$-gerbes on $B\times S^1.$ If the characteristic class of
the bundle $E$ is $[E]$, the $H$-flux of the T-dual is given by $[E]
\times z$, where $z$ is the canonical generator of $H^1(S^1)$.

We would like to extend this correspondence to semi-free
$S^1$-actions.
A semi-free space $X$ is completely specified by the fixed point set
$F \subset B$ and the class of the principal $S^1$-bundle over $B-F$.
Thus, it is specified by a class $\lambda$ in $H^2(B-F)$.

This class may be used to construct a $2$-gerbe on $B^{+}$ by taking
the image of $\lambda$ in $H^3(B^{+})$ by the following
sequence\footnote{This sequence is not exact at $H^3(B,B-F)$
otherwise this class would always\break be~zero.} (see the previous
section):
\begin{equation}
H^2(B-F) \longrightarrow H^3(B,B-F) \longrightarrow H^3(B^{+},B^{+}-F) \longrightarrow H^3(B^{+}).
\label{LESBF}
\end{equation}

The T-dual of such a space is, by the argument presented in
Chapter~1, the space $B \times S^1$ with a source of $H$-flux located
at $F \times S^1$. Such a source emits a $H$-flux which defines a
class $\lambda \times z$ in $H^3((B-F)\times S^1)$.

This class may be used to construct a $3$-gerbe on $B^{+} \times S^1$
by taking the image of $\lambda \times z$ in $H^3(B^{+})$ via the
following sequence\footnote{This sequence too is not exact at
$H^3(B,B-F)$ otherwise this class would always be~zero.} (see the
previous section):
\begin{align} \label{LESBFS1}
H^3((B-F)\times S^1)&\longrightarrow  H^4(B \times S^1,(B-F)\times S^1) \nonumber \\
&\longrightarrow  H^4(B^{+} \times S^1, (B^{+} - F) \times S^1) \longrightarrow  H^4(B^{+}
\times S^1).
\end{align}

Thus, there seems to be a map between $2$-gerbes on $B^{+}$ and
$3$-gerbes on $B^{+}\times S^1$ induced by T-duality. This may be
understood as follows. If we fix a generator $z$ of $H^1(S^1)$, then
taking the cross product of a cohomology class $\lambda \in H^k(X)$
with $z$ gives a homomorphism $\times :H^k(X) \to H^{k+1}(X \times
S^1)$. If $k=2$, this homomorphism is exactly the one which is
induced by sending the characteristic class of a principal $S^1$
bundle to the H-flux on the T-dual trivial bundle. It is interesting
therefore, that the same map for $k=3$ is also induced by T-duality.

\setcounter{theorem}{2}
\begin{theorem}
Let ${\mathcal \mathbf X}$ be a $2$-gerbe on $B^{+}$ with
characteristic class $\eta \in H^3(B^{+})$. T-duality defines a map
which sends ${\mathcal \mathbf X}$ to a $3$-gerbe ${\mathcal \mathbf
Y}$ on $B^{+} \times S^1$ with characteristic class $(\eta \times z)
\in H^4(B^{+} \times S^1)$. \label{ThmGerbe}
\end{theorem}

\begin{proof}
Pick an open cover $\{ U_i \}$ for $B^{+}.$ Then, the $2$-gerbe on
$B^{+}$ induces principal $S^1$-bundles $p_{ij}:L^{j}_{i} \to
U_{ij}$. Let $[p_{ij}]$ denote the characteristic class of
$L^{i}_{j}$. By the definition of a 2-gerbe (see
Definition~\ref{2GerbDef}.), on $U_{ijk}$ the bundle
\begin{equation}
L^j_i\vert_{U_{ijk}}\otimes L^k_j\vert_{U_{ijk}}\otimes
L^i_k\vert_{U_{ijk}} \label{EqVbTen}
\end{equation}
is trivial with a canonical section $\theta_{ijk}$. The definition
requires $\delta \theta = 1$. We take the cohomology class of
$\theta$ to be $\eta.$ T-dualizing each of the bundles $L^i_j$ gives
continuous-trace algebras $A^i_j$ on $U_{ij} \times S^1$ with
characteristic class $[p_{ij}] \times z$. Note that the
characteristic class of $A^i_j$ and $A^j_i$ are inverses of each
other in $H^3(U_{ij})$ since
\[([p_{ij}] \times z) + ([p_{ji}] \times
z) = ([p_{ij}]+[p_{ji}])\times z = 0.\] Let $w_{\alpha}: U_{ijk} \to
U_{\alpha}, \alpha = ij,jk,ki$ denote the inclusion map. Then, since
the tensor product in equation~(\ref{EqVbTen}) is trivial, we see
that
\begin{equation}
w_{ij}^{\ast}([p_{ij}]) + w_{jk}^{\ast}([p_{jk}]) +
w_{ki}^{\ast}([p_{ki}]) =0. \label{EqCCLineb}
\end{equation}
Let us try to compute the characteristic class of the tensor
product
\begin{equation}
{\mathcal A}^{j}_{i}\vert_{U_{ijk}} \otimes {\mathcal A}^{k}_{j}
\vert_{U_{ijk}} \otimes {\mathcal A}^{i}_{k} \vert_{U_{ijk}}.
\label{EqCTAlgTen}
\end{equation}
This would be given by
\[
(w_{ij} \times 1)^{\ast}([p_{ij}] \times z)
+ (w_{jk} \times 1)^{\ast}([p_{jk}] \times z) + (w_{ki} \times
1)^{\ast}([p_{ki}] \times z),
\]
where
\[
w_{\alpha} \times 1: (U_{ijk}\times S^1) \longrightarrow (U_{\alpha} \times S^1),
\alpha = \{ij, jk, ki\},
\]
are the induced inclusion maps on the T-dual side. This may be
simplified as follows
\begin{align}
&(w_{ij} \times 1)^{\ast}([p_{ij}] \times z) + (w_{jk} \times
1)^{\ast}([p_{jk}] \times z) +
(w_{ki} \times 1)^{\ast}([p_{ki}] \times z) \nonumber \\
&\quad= w_{ij}^{\ast}([p_{ij}]) \times z +  w_{jk}^{\ast}([p_{jk}])
\times z
+ w_{ki}^{\ast}([p_{ki}]) \times z \nonumber\\
&\quad= (w_{ij}^{\ast}([p_{ij}]) +  w_{jk}^{\ast}([p_{jk}]) +
w_{ki}^{\ast}([p_{ki}])) \times z  = 0. \label{EqCTAUijk}
\end{align}

Thus, the continuous-trace algebra defined in
equation~(\ref{EqCTAlgTen}) is trivial. Thus, it must possess a
section \footnote{See Definition~\ref{3GerbDef}.} which would be a
line bundle $\Gamma_{ijk}$ over $U_{ijk} \times S^1$. To obtain this
section, we note that $\theta_{ijk}$ defines an element $[
\theta_{ijk} ] \in H^1(U_{ijk};\KZ)$ and so we obtain an element $([
\theta_{ijk}] \times z) \in H^2(U_{ijk} \times S^1; \KZ)$ which
defines $\Gamma_{ijk}$.

Now, by Definition~(\ref{3GerbDef}), restricting these $\Gamma$ to
$U_{ijkl}$ and calculating the tensor product
\begin{equation}
\Gamma_{ijk}\vert_{U_{ijkl}}\otimes
\Gamma_{ijl}^{-1}\vert_{U_{ijkl}} \otimes
\Gamma_{ikl}\vert_{U_{ijkl}} \otimes
\Gamma_{jkl}^{-1}\vert_{U_{ijkl}} \label{EqGammaTen}
\end{equation}
should give us a trivial bundle and a canonical section $\eta_{ijkl}$
which is a Cech cocycle. To show that the tensor product
equation~(\ref{EqGammaTen}) is trivial, we once again calculate the
characteristic class of this tensor product line bundle. If
$w_{\alpha}:U_{ijkl} \to U_{\alpha}, \alpha=\{ijk,ijl,ikl,jkl\}$ is
the inclusion map, the class we want to calculate is
\begin{align*}
&(w_{ijk}\times 1)^{\ast}([\theta_{ijk}]\times z) -(w_{ijl} \times
1)^{\ast}([\theta_{ijl}]\times z) +(w_{ikl} \times
1)^{\ast}([\theta_{ikl}]\times z) \\
&\qquad -(w_{jkl} \times 1)^{\ast}([\theta_{jkl}]\times z)
\end{align*}
This may be simplified as follows
\begin{align}
&(w_{ijk} \times 1)^{\ast}([\theta_{ijk}]\times z)
-(w_{ijl} \times 1)^{\ast}([\theta_{ijl}]\times z) \nonumber \\
&\qquad+(w_{ikl} \times 1)^{\ast}([\theta_{ikl}]\times z)
-(w_{jkl} \times 1)^{\ast}([\theta_{jkl}]\times z) \nonumber \\
&\quad= (w_{ijk}^{\ast}([\theta_{ijk}])
-w_{ijl}^{\ast}([\theta_{ijl}]) +w_{ikl}^{\ast}([\theta_{ikl}])
-w_{jkl}^{\ast}([\theta_{jkl}]) )\times z. \label{EqGammaUijkl}
\end{align}
The term in parenthesis in the last equation is the class in
$H^1(U_{ijkl})$ induced by $\delta \theta.$ Since $\delta \theta=1$,
the expression vanishes. Note that if we change $\theta$ by a
coboundary, the $\Gamma$ will change, but the tensor product will
still remain trivial as its characteristic class will shift by the
class in $H^1(U_{ijkl})$ of the coboundary of a coboundary.

We now need a trivialization of this tensor product on 5-fold
intersections. This is given by any representative of the cross
product cocycle $\theta \times z$ which gives a cocycle on 5-fold
intersections and so a $\KC^{\ast}$-valued function on this space.
Changing the cocycle within its cohomology class will not change the
gerbe as the characteristic class of the gerbe will remain the same.

Changing the original cover $\{U_i\}$ will not affect the answer, as
following the above construction through on the new cover will show.
It is also clear that the characteristic class of the $3$-gerbe so
constructed will be $\eta \times z$.
\end{proof}

Now, if we are given a space with a semi-free $S^1$-action with fixed
point set whose image is $F \subset B$, then, as argued above, we get
a class in $H^2(B-F)$ and a gerbe on $B^{+}$. If we pick an open
cover of $B^{+}$ containing $B-F$ and $B-\{ + \}$, under T-duality,
we will obtain by the above theorem a $3$-gerbe on $B^{+} \times S^1$
whose restriction to $B-F$ is exactly the continuous-trace algebra
which is the T-dual of the line bundle we had over $B-F$.

We saw above that a semi-free $S^1$-space $X$ with quotient space
a\break (compact, closed, and connected) manifold $B$ is classified up to
equivariant homeomorphism by the fixed point set $F \subset B$ and
the characteristic class of the principal $S^1$-bundle
($X-\pi^{-1}(F)) \overset{\pi}{\to} (B-F)$. We now assume that $F$ is
a smooth embedded submanifold of $B$. We associated to $X$ a
cohomology class in $H^3(B,B-F)$ which gave us a class in
$H^3(B^{+},B^{+}-F)$ by the excision isomorphism and finally gave us
a class in $H^3(B^{+})$ (using the long exact sequence of the pair
$(B^{+},B^{+}-F)$). However, we could have obtained a class in
$H^3(\tilde{B})$ for any compactification $\tilde{B}$ of $B$.
($B^{+}$ is not always a manifold even if $B$ is, so in applications
we might need to use another compactification $\tilde{B}$.)

\setcounter{lemma}{3}
\begin{lemma}
There is a space $Y,$ a map $H^3(B,B-F)\to H^3(Y)$ together with a
natural map $\phi: H^3(Y) \to H^3(\tilde{B})$ such that every map
$H^3(B,B-F) \to H^3(\tilde{B})$ factors through $\phi$.
\end{lemma}

\begin{proof}
This space may be constructed as follows. If $N(F)$ is a tubular
neighborhood of $F$ in $B,$ then, by the tubular neighborhood
theorem, $N(F)$ is diffeomorphic to the normal bundle of $F$ in $B.$
Let $D(F)$ be the closure of $N(F)$ in $B,$ then, $D(F)$ is
homeomorphic to a disc bundle over $F.$ By excision, and homotopy,
\[
H^{k}(B,B-F) \simeq H^{k}(D(F),D(F)-F) \simeq H^{k}(D(F),S(F)),
\]
where $S(F)$ is the sphere bundle which is the boundary of the disc
bundle $D(F)$. Now, $H^k(D(F),S(F)) \simeq H^k(D(F)/S(F)) \simeq
H^k(\hbox{TD}(F))$ where $\hbox{TD}(F)$ is the Thom space of the disc bundle
$D(F)$ (see Ref. \cite[page 441 for details]{ATop}). Now, for any
space $\tilde{B}$ containing $F,$ there is a collapse map $\lambda:
\tilde{B} \to \hbox{TD}(F)$ obtained by collapsing everything outside
$N(F) \subset B \subset \tilde{B}$ to a point.  The following diagram
commutes:
\begin{equation}
\CD \minCDarrowwidth1pc
H^3(D(F),S(F))@>{\simeq}>> H^3(\hbox{TD}(F),{\ast}) @>>> H^3(\hbox{TD}(F)) \\
@V{\simeq}VV  @. @V{\lambda^{\ast}}VV \\
H^3(B,B-F) @>{\simeq}>> H^3(\tilde{B},\tilde{B}-F) @>>>
H^3(\tilde{B})
\endCD
\label{CDThom1}
\end{equation}

>From this it follows that the image of any class in $H^3(\tilde{B})$
which is the image of a class $\gamma$ in $H^3(B,B-F)$ is actually
pulled back from the image of $\gamma$ in  $H^3(\hbox{TD}(F))$ via
$\lambda^{\ast}.$ Thus, the image of $\gamma$ in $H^3(\hbox{TD}(F))$
is a {\em universal} invariant.
\end{proof}

It follows from this construction that the $2$-gerbe we constructed
in the previous subsection is actually pulled back from the $2$-gerbe
on $Y$ via the collapse map. It also follows from this construction
that the invariant is zero once the codimension $k$ of $F$ is more
than $3.$ For, by a property of the Thom space (See Ref. \cite[page
441]{ATop}), $H^i(\hbox{TD}(F)) = 0$ if $i < k.$ Further, we have the
Thom isomorphism $\Phi: H^i(F) \to H^{i+k}(\hbox{TD}(F))$. This
should enable us to calculate the invariant explicitly.

On the T-dual side, we have a trivial $S^1$-bundle $B \times S^1$
with the NS$5$-brane sitting somewhere in $F \times S^1,$ transverse
to the $S^1$-fiber. This would have a total charge given by a
cohomology class in $H^4((B\times S^1),(B-F)\times S^1)$. By an
argument similar to the above, we would have a commutative diagram.
{\fontsize{9.5}{11}\selectfont\begin{gather} \xymatrix{
 & {H^3(D(F \times S^1), S(F \times S^1)) } \ar[dr]_{\simeq}
\ar[dl]_{\simeq} & \\
 {H^3(B \times S^1, (B-F) \times S^1)} \ar[d]^{\simeq} &
& { H^3(F \times S^1, \ast)} \ar[d]\\
{H^3(\tilde{B} \times S^1, (\tilde{B} - F) \times S^1)} \ar[dr]
& &
H^3(\hbox{TD}(F \times S^1)) \ar[dl]^{\lambda^{\ast}}\\
& {H^3(\tilde{B} \times S^1)} & \\
} \label{CDThom2}
\end{gather}}

We also have a Thom isomorphism $\tilde{\Phi}: H^i(F\times S^1)
\!\to\! H^{i+k}(\hbox{TD}(F\times S^1))$. Recall that T-duality gave
a map $\times:H^3(B-F) \to H^4((B-F)\times S^1)$ given by the cross
product with $z \in H^1(S^1)$. We saw in the previous section that
this induced a map $\times:H^3(B^{+}) \to H^3(B^{+} \times S^1)$. An
argument similar to the one given in that section would give also
give a map $\times:H^3(\tilde{B}) \to H^4(\tilde{B} \times S^1)$.
Then we have a commutative diagram
\begin{equation}
\CD \minCDarrowwidth1pc H^{4-k}(F \times S^1) @>{\Phi^{-1}}>>
H^4(\hbox{TD}(F \times S^1))
@>{\lambda^{\ast}}>> H^4(\tilde{B} \times S^1) \\
@A{\times}AA @. @A{\times}AA \\
H^{3-k}(F) @>{\Phi^{-1}}>> H^3(\hbox{TD}(F)) @>{\lambda^{\ast}}>>
  H^3(\tilde{B}) \\
\endCD
\label{CDThom3}
\end{equation}

This may be used to calculate the invariant in $H^4$ for an
NS$5$-brane\break configuration from the one in $H^3$.

\section{Conclusions and future work}

In this paper, we have studied the Topological T-dual of a semi-free
$S^1$-space. We have shown how to model a wrapped NS5-brane within
the formalism of Topological T-duality as an extension of
$C^{\ast}$-algebras. We justify this by showing how it may be used to
model the dyonic coordinate of a\break KK-monopole. We also study this
extension using gerbes. We summarize the contents of this paper by
the following commutative diagram. Each of the horizontal arrows is
one of the results of this paper:
\begin{gather}
\xymatrix{ { \txt{ (2-Gerbe on $B^+$)} } \ar@{<=>}[rr]^{\txt{Theorem.
2.2}} &&
{ \txt{(3-Gerbe on $B^+ \times S^1$)}} \\
{ \left( \txt{ Semi-free space $X$ \\ with $X/S^1 \simeq B$ }
\right) } \ar@{<=>}[rr]^{\txt{T-duality}} \ar@{=>}[u]  &&
{ \left( \txt{ $B \times S^1$ with \\ source of $H$-flux \\
on $F \times S^1$ } \right)  }
\ar@{=>}[u] \\
{C_0(X,\Cpct) } \ar@{<=>}[rr]^{\txt{Crossed \\ Product}}
\ar@{<=>}[u] &&
{\left( \txt{ Extension \\ like equation~(\ref{SES1}). } \right)} \ar@{<=>}[u] \\
} \label{CDFinal}
\end{gather}

It would be interesting to extend this to semi-free $T^n$ spaces. It
is shown in Ref.~\cite{MRCMP} that it may not always be possible to
T-dualize such a space. However, it would be interesting to see if we
could study the ``brane box" \cite{BBox} configurations using some
extension of the present formalism.

We showed using topological T-duality in Section~\ref{SecKKTdual}
that the algebra obtained as the extension equation~(\ref{SES1}) is
entirely determined by its spectrum, a cohomology class in $(X-F)/S^1
\times S^1$, together with the fixed point set $F \subset X/S^1$. Is
there an alternative characterisation of this $C^{\ast}$-algebra,
independent of Topological T-duality? If so, this algebra would be a
model for spaces with an arbitrary configuration of NS5-branes.

\enlargethispage{10pt}

\section*{Acknowledgments}

I thank Professor J. Rosenberg, University of Maryland, College Park,
for key insights and helpful discussions. I also gratefully
acknowledge the\break Research Assistantship which made this work
possible. This Assistantship was supported by the NSF grant
DMS-0504212.

\end{document}